\RequirePackage{fix-cm} 
\RequirePackage[reqno]{amsmath}
\RequirePackage[l2tabu, orthodox]{nag}
\documentclass[a4paper, twoside, dvips, reqno, 12pt]{amsart}
\usepackage{etex}
\usepackage{fixltx2e}   

\usepackage[latin1]{inputenc}
\usepackage[T1]{fontenc}

\usepackage{esint}
\usepackage{amsbsy}
\usepackage{dsfont}
\usepackage{empheq}
\usepackage{xspace}
\usepackage{amsgen}
\usepackage{amsthm}
\usepackage{manfnt}
\usepackage{pifont}
\usepackage{amssymb}
\usepackage{upgreek}
\usepackage{amsfonts}
\usepackage{extarrows}
\usepackage{mathtools}
\usepackage{textgreek}
\usepackage[nice]{nicefrac}

\usepackage{mathabx}
\usepackage{relsize}
\usepackage{textcomp}
\usepackage[mathcal]{euscript}

\usepackage{rsfso}
\usepackage{mathrsfs}
\DeclareMathAlphabet{\mathscrbf}{OMS}{mdugm}{b}{n}

\DeclareFontFamily{U}{dutchcal}{\skewchar\font=45 }
\DeclareFontShape{U}{dutchcal}{m}{n}{<-> s*[1.0] dutchcal-r}{}
\DeclareFontShape{U}{dutchcal}{b}{n}{<-> s*[1.0] dutchcal-b}{}
\DeclareMathAlphabet{\mathlcal}{U}{dutchcal}{m}{n}
\SetMathAlphabet{\mathlcal}{bold}{U}{dutchcal}{b}{n}

\usepackage{a4wide}

\usepackage[dvipsnames, table]{xcolor}
\usepackage[most]{tcolorbox}

\definecolor{bckg}{RGB}{20.8, 20.8, 20.8}
\definecolor{oneblue}{rgb}{0.0, 0.0, 0.85}
\definecolor{Lightblue}{RGB}{214, 214, 214}
\definecolor{bluepigment}{rgb}{0.2, 0.2, 0.6}
\definecolor{charcoal}{rgb}{0.21, 0.27, 0.31}
\definecolor{denimblue}{rgb}{0.08, 0.38, 0.74}
\definecolor{Lightgray}{rgb}{0.89, 0.89, 0.89}
\definecolor{darkgrey}{rgb}{0.273, 0.281, 0.30}
\definecolor{darkelectricblue}{rgb}{0.33, 0.41, 0.47}

\usepackage[sort&compress, comma, square, numbers]{natbib}

\usepackage{tikz}
\usepackage{psfrag}
\usepackage{graphicx}
\usepackage{subfigure}
\usepackage{morefloats}
\usepackage{indentfirst}

\usepackage{acronym}
\usepackage{microtype}
\usepackage[labelsep=period,%
            labelfont={bf,sf,color=bluepigment},%
            justification=raggedright]{caption}

\usepackage[perpage, symbol]{footmisc}

\usepackage[usenames, dvipsnames, pdf]{pstricks}
\usepackage{pst-grad} 
\usepackage{pst-plot} 

\usepackage[colorlinks,
           urlcolor=oneblue,
           linkcolor=denimblue,
           citecolor=NavyBlue,
           bookmarksopen=false,
           pdfpagemode=UseNone,
           plainpages=false,
           pagebackref]{hyperref}

\usepackage[explicit]{titlesec}

\titleformat{\section}
  {\color{NavyBlue}\Large\sffamily\bfseries}
  {}
  {0em}
  {\colorbox{bckg!5}{\parbox{\dimexpr\linewidth-2\fboxsep\relax}{\centering\thesection. #1}}}
  [\vspace*{0.33em}]

\titleformat{name=\section, numberless}
  {\color{NavyBlue}\Large\sffamily\bfseries}
  {}
  {0.0em}
  {\colorbox{bckg!5}{\parbox{\dimexpr\linewidth-2\fboxsep\relax}{\centering#1}}}
  [\vspace*{0.33em}]

\titleformat{\subsection}
  {\color{NavyBlue}\large\sffamily\bfseries}
  {}
  {0.0em}
  {\colorbox{bckg!5}{\parbox{\dimexpr\linewidth-5\fboxsep\relax}{\centering\thesubsection. #1}}}
  [\vspace*{0.33em}]

\titleformat{name=\subsection, numberless}
  {\color{NavyBlue}\Large\sffamily\bfseries}
  {}
  {0em}
  {\colorbox{bckg!5}{\parbox{\dimexpr\linewidth-5\fboxsep\relax}{\centering#1}}}
  [\vspace*{0.33em}]

\titleformat{\subsubsection}
  {\color{bluepigment}\sffamily\normalsize\bfseries}
  {}
  {0.0em}
  {\colorbox{bckg!5}{\parbox{\dimexpr\linewidth-5\fboxsep\relax}{\centering\thesubsubsection. #1}}}
  [\vspace*{0.33em}]

\titleformat{name=\subsubsection, numberless}
  {\color{bluepigment}\sffamily\normalsize\bfseries}
  {}
  {0.0em}
  {\colorbox{bckg!5}{\parbox{\dimexpr\linewidth-5\fboxsep\relax}{\centering#1}}}
  [\vspace*{0.33em}]

\titleformat{\paragraph}[runin]
  {\color{bluepigment}\sffamily\small\bfseries}
  {}
  {0em}
  {#1}

\titlespacing{\section}{1.0em}{1.5em plus 2pt minus 2pt}%
{1.0em plus 2pt minus 2pt}[0em]
\titlespacing{\subsection}{1.0em}{1.5em plus 2pt minus 2pt}%
{1.0em}[0em]
\titlespacing{\subsubsection}{1.0em}{1.5em plus 2pt minus 2pt}%
{1.0em plus 2pt minus 2pt}[0em]

\usepackage{titletoc}

\contentsmargin{0.5em}
\newlength{\tocsep} 
\titlecontents{section}[\tocsep]
  {\addvspace{10pt}\bfseries\sffamily}
  {\contentslabel[\thecontentslabel]{\tocsep}}
  {}
  {\ \titlerule*[0.75pc]{.}\ \thecontentspage}
  []

\titlecontents{subsection}[\tocsep]
  {\addvspace{8pt}\sffamily}
  {\contentslabel[\thecontentslabel]{\tocsep}}
  {}
  {\ \titlerule*[0.5pc]{.}\ \thecontentspage}
  []

\titlecontents{subsubsection}[\tocsep]
  {\addvspace{4pt}\footnotesize\sffamily}
  {\contentslabel[\thecontentslabel]{\tocsep}\hspace*{0.5em}}
  {}
  {\ \titlerule*[0.35pc]{.}\ \thecontentspage}
  []

\makeatletter
\def\@setauthors{%
  \begingroup
  \def\thanks{\protect\thanks@warning}%
  \trivlist
  \centering\footnotesize \@topsep30\p@\relax
  \advance\@topsep by -\baselineskip
  \item\relax
  \author@andify\authors
  \def\\{\protect\linebreak}%
  \textsc{\normalsize\textcolor{darkelectricblue}{\authors}}%
  \ifx\@empty\contribs
  \else
    ,\penalty-3 \space \@setcontribs
    \@closetoccontribs
  \fi
  \endtrivlist
  \endgroup
}
\def\@settitle{\begin{center}%
  \baselineskip14\p@\relax
    \bfseries
    \textsc{\Large\textcolor{charcoal}{\@title}}
  \end{center}%
}
\makeatother

\usepackage{enumitem}
\setlist[description]{%
  topsep = 30pt,               
  itemsep = 8pt,               
  labelsep = 10pt,
  font={\bfseries\color{NavyBlue}}, 
}

\usepackage{fancyhdr}
\usepackage{lastpage}

\newcommand*\Title{\textcolor{bluepigment}{Non-dispersive regularization of shallow-water equations}}
\newcommand*\Authors{\textcolor{bluepigment}{Y.~Pu, R.~Pego, D.~Dutykh \& D.~Clamond}}
\newcommand*{\plogo}{\textcolor{gray}{{\texttt{arXiv.org} / \textsc{hal}}}} 

\numberwithin{equation}{section}

\theoremstyle{plain}

\theoremstyle{remark}

\newtcbox{\mymath}[1][]{%
    nobeforeafter, math upper, tcbox raise base,
    enhanced, colframe = black!35,
    colback = black!5, boxrule = 1pt, arc = 0mm,
    #1}

\newcommand{\hnon}{H}
\newcommand{\vnon}{V}
\newcommand{\D}{\partial}
\newcommand{\inv}{^{\,-1}}
\newcommand{\R}{\mathds{R}}
\newcommand{\ud}{\mathrm{d}}
\newcommand{\ui}{\mathrm{i}}
\newcommand{\ue}{\mathrm{e}}
\newcommand{\eeo}{\eee^{\,0}}
\newcommand{\qqo}{\qqq^{\,0}}
\newcommand{\eps}{\varepsilon}
\newcommand{\eep}{\eee^{\,\eps}}
\newcommand{\qqp}{\qqq^{\,\eps}}
\newcommand{\ddd}{{\mathcal{D}}}
\newcommand{\eee}{{\mathcal{E}}}
\newcommand{\fff}{{\mathcal{F}}}
\newcommand{\rrr}{{\mathcal{R}}}
\newcommand{\qqq}{{\mathcal{Q}}}
\newcommand{\calH}{{\mathcal H}}
\newcommand{\calR}{{\mathcal R}}
\newcommand{\llll}{{\mathcal{L}}}
\newcommand{\half}[1][1]{{\frac{#1}{2}}}
\newcommand{\paren}[1]{{\left(#1\right)}}
\newcommand{\bracket}[1]{{\left[#1\right]}}
\newcommand{\eqdef}{\stackrel{\textup{def}}{:=}}

\renewcommand{\O}{\mathcal{O}}
\newcommand{\RR}{{\mathds{R}}}
\newcommand{\abs}[1]{\lvert\, #1\, \rvert}
\newcommand{\angl}[1]{{\left\langle\;#1\;\right\rangle}}

\DeclareMathOperator*{\supp}{supp}
\newcommand{\sech}{\operatorname{sech}}

\begin{document}

\title[\Title]{Weakly singular shock profiles for a non-dispersive regularization of shallow-water equations}

\author[Y.~Pu]{Yue~Pu}
\address{\textbf{Y.~Pu:} Department of Mathematical Sciences and Center for Nonlinear Analysis, Carnegie Mellon University, Pittsburgh, Pennsylvania, PA 12513, USA}
\email{ypu@andrew.cmu.edu}

\author[R.~L.~Pego]{Robert~L.~Pego$^*$}
\address{\textbf{R.~L.~Pego:} Department of Mathematical Sciences and Center for Nonlinear Analysis, Carnegie Mellon University, Pittsburgh, Pennsylvania, PA 12513, USA}
\email{rpego@cmu.edu}
\urladdr{http://www.math.cmu.edu/~bobpego/}
\thanks{$^*$ Corresponding author}

\author[D.~Dutykh]{Denys Dutykh}
\address{\textbf{D.~Dutykh:} Univ. Grenoble Alpes, Univ. Savoie Mont Blanc, CNRS, LAMA, 73000 Chamb\'ery, France and LAMA, UMR 5127 CNRS, Universit\'e Savoie Mont Blanc, Campus Scientifique, 73376 Le Bourget-du-Lac Cedex, France}
\email{Denys.Dutykh@univ-smb.fr}
\urladdr{http://www.denys-dutykh.com/}

\author[D.~Clamond]{Didier Clamond}
\address{\textbf{D.~Clamond:} Universit\'e C\^ote d'Azur, Laboratoire J. A. Dieudonn\'e, CNRS UMR 7351, Parc Valrose, F-06108 Nice cedex 2, France}
\email{diderc@unice.fr}
\urladdr{http://math.unice.fr/~didierc/}


\begin{titlepage}
\clearpage
\pagenumbering{arabic}
\thispagestyle{empty} 
\noindent
{\Large Yue \textsc{Pu}}\\
{\textit{\textcolor{gray}{Carnegie Mellon University, USA}}}
\\[0.02\textheight]
{\Large Robert~L.~\textsc{Pego}}\\
{\textit{\textcolor{gray}{Carnegie Mellon University, USA}}}
\\[0.02\textheight]
{\Large Denys \textsc{Dutykh}}\\
{\textit{\textcolor{gray}{CNRS, Universit\'e Savoie Mont Blanc, France}}}
\\[0.02\textheight]
{\Large Didier \textsc{Clamond}}\\
{\textit{\textcolor{gray}{Universit\'e C\^ote d'Azur, France}}}
\\[0.16\textheight]

\colorbox{Lightblue}{
  \parbox[t]{1.0\textwidth}{
    \centering\huge
    \vspace*{0.75cm}
    
    \textsc{\textcolor{bluepigment}{Weakly singular shock profiles for a non-dispersive regularization of shallow-water equations}}
    
    \vspace*{0.75cm}
  }
}

\vfill 

\raggedleft     
{\large \plogo} 
\end{titlepage}


\cleardoublepage
\thispagestyle{empty} 
\par\vspace*{\fill}   
\begin{flushright} 
{\textcolor{denimblue}{\textsc{Last modified:}} \today}
\end{flushright}


\clearpage
\maketitle
\thispagestyle{empty}


\begin{abstract}

We study a regularization of the classical \textsc{Saint--Venant} (shallow-water) equations, recently introduced by D.~\textsc{Clamond} and D.~\textsc{Dutykh} ({\it Commun. Nonl. Sci. Numer. Simulat.} \textbf{55} (2018) 237--247). This regularization is non-dispersive and formally conserves mass, momentum and energy. We show that for every classical shock wave, the system admits a corresponding non-oscillatory traveling wave solution which is continuous and piecewise smooth, having a weak singularity at a single point where energy is dissipated as it is for the classical shock. The system also admits cusped solitary waves of both elevation and depression.


\bigskip
\noindent \textbf{\keywordsname:} \textsc{Serre} equations, \textsc{Green--Naghdi} equations, shallow water, weak solutions, long waves, peakons, cuspons, energy loss. \\


\end{abstract}


\newpage
\tableofcontents
\thispagestyle{empty}


\newpage
\section{Introduction}

In a recent paper, \textsc{Clamond} and \textsc{Dutykh} \cite{Clamond2018} have introduced a regularization of the classical \textsc{Saint-Venant} (shallow-water) equations, which is non-dispersive, non-dissipative, and formally conserves mass, momentum, and energy. In conservation form, these regularized \textsc{Saint-Venant} equations (rSV) are written
\begin{gather}
  h_{\,t}\ +\ (h\,u)_{\,x}\ =\ 0 \,,\label{e:rsvh} \\ 
  (h\,u)_{\,t}\ +\ (h\,u^{\,2}\ +\ \frac12\;g\,h^{\,2}\ +\ \eps\,\calR\,h^{\,2})_{\,x}\ =\ 0\,, \label{e:rsvu} \\
  \calR\ \eqdef\ h\,(u_{\,x}^{\,2}\ -\ u_{\,x\,t}\ -\ u\,u_{\,x\,x})\ -\ g\,\left(h\,h_{\,x\,x}\ +\ \frac12\;h_{\,x}^{\,2}\right)\,.\label{e:rsvR}
\end{gather}
Smooth solutions of these equations also satisfy a conservation law for energy, in the form
\begin{equation}\label{e:rsvE}
  \eep_{\,t}\ +\ \qqp_{\,x}\ =\ 0\,,
\end{equation}
where
\begin{align}\label{d:eep}
  & \eep\ \eqdef\ \half\;h\,u^{\,2}\ +\ \half\;g\,h^{\,2}\ +\ \eps\,\left(\half\;h^{\,3}\,u_{\,x}^{\,2}\ +\ \half\;g\,h^{\,2}\,h_{\,x}^{\,2}\right)\,, \\
  & \qqp\ \eqdef\ \half\;h\,u^{\,3}\ +\ g\,h^{\,2}\,u\ +\ \eps\,\left(\paren{\half\;h^{\,2}\,u_{\,x}^{\,2}\ +\ \half\;g\,h\,h_{\,x}^{\,2}\ +\ h\,\calR}\,h\,u\ +\ g\,h^{\,3}\,h_{\,x}\,u_{\,x}\right)\,.
\end{align}
The rSV equations \eqref{e:rsvh} -- \eqref{e:rsvu} above were derived in \cite{Clamond2018} as the \textsc{Euler--Lagrange} equations corresponding to a least action principle for a \textsc{Lagrangian} of the form (see \cite[Eq.~(3.2)]{Clamond2018})
\begin{equation*}
  \llll\ \eqdef\ \half\;h\,u^{\,2}\ -\ \half\;g\,h^{\,2}\ +\ \eps\,\left(\half\;h^{\,3}\,u_{\,x}^{\,2}\ -\ \half\;g\,h^{\,2}\,h_{\,x}^{\,2}\right)\ +\ \paren{h_{\,t}\ +\ (h\,u)_{\,x}}\;\phi\,.
\end{equation*}
Here $\phi$ is a \textsc{Lagrange} multiplier field that enforces mass conservation. The terms proportional to $\eps$ in \eqref{e:rsvu} have a form similar to terms that appear in improved \textsc{Green}--\textsc{Naghdi} or \textsc{Serre} equations that approximate shallow-water dynamics for waves of small slopes, see \cite{Clamond2015c}. (The rSV equations also admit a non-canonical \textsc{Hamiltonian} structure like one known for the \textsc{Green}--\textsc{Naghdi} equations --- see Section~\ref{sec:6} below.)

The particular coefficients appearing here, however, do not yield improved accuracy for modeling exact water-wave dispersion at long wavelengths. Instead, they are designed to \emph{eliminate} linear dispersion, resulting in a regularization that faithfully reproduces the original shallow-water dispersion relation. The balance of terms in $\calR$ ensures that the rSV equations are \emph{non-dispersive} --- linearized about a constant state $(h_{\,0},\,u_{\,0})\,$, solutions proportional to $\ue^{\,\ui\,k\,x\ -\ \ui\,\omega\,t}$ necessarily have 
\begin{equation*}
  (\omega\ -\ u_{\,0}\,k)^{\,2}\ =\ g\,h_{\,0}\,k^{\,2}\,,
\end{equation*}
implying that phase velocity is independent of frequency.

The presence of squared derivatives in the energy $\eep$ indicates that the rSV equations will not admit classical shock wave solutions with discontinuities in $h$ and $u\,$. Numerical experiments reported in \cite{Clamond2018} suggest, in fact, that with smooth initial data that produce hydraulic jumps (shock wave solutions) for the shallow-water equations, one obtains front-like solutions of the rSV equations that remain smooth and non-oscillatory, yet propagate at the correct speed determined by classical jump conditions corresponding to limiting states on the left and right. These solutions were computed numerically by a pseudospectral scheme that is highly accurate for smooth solutions and fairly uncomplicated. This is a hint that a similar approach could perhaps be taken to approximate shallow water dynamics by non-dispersive regularization in multidimensional geometries with more complicated topography and other physics.

At this point, a paradox arises. The energy of smooth solutions of the rSV equations satisfies the conservation law~\eqref{e:rsvE}, whereas in the case of shallow water equations, energy is dissipated at a shock-wave discontinuity, satisfying a distributional identity of the form
\begin{equation}\label{e:eeo}
  \eeo_{\,t}\ +\ \qqo_{\,x}\ =\ \mu\,,
\end{equation}
where $\mu$ is a non-positive measure supported along the shock curve. How can it be that front-like solutions of the rSV equations approximate classical shallow-water shocks well while conserving an energy similar to the one dissipated for shallow-water shocks?

Our purpose here is to describe a novel wave-propagation mechanism that may explain this paradox. We shall show that the regularized \textsc{Saint-Venant} equations \eqref{e:rsvh} -- \eqref{e:rsvu} admit regularized shock-wave solutions with profiles that are \emph{continuous but only piecewise smooth}, with derivatives having a weak singularity at a single point. Such a wave exists corresponding to every classical shallow-water shock. These waves are traveling-wave weak solutions of the rSV equations that conserve mass and momentum. They \emph{dissipate energy at the singular point}, however, at the precise rate that the corresponding classical shock does.

We also find that the rSV equations admit weak solutions in the form of \emph{cusped solitary waves}. These waves loosely resemble the famous `peakon' solutions of the \textsc{Camassa--Holm} equation in the non-dispersive case \cite{Camassa1993}. One difference is that the wave slope of our cusped solitary waves becomes infinite approaching the crest, while that of a peakon remains finite. The rSV equations also loosely resemble various $2-$component generalizations of the \textsc{Camassa--Holm} equation which have appeared in the literature---for a sample see \cite{Chen2006b, Ivanov2006, Kuzmin2007, Holm2009, Ionescu-Kruse2013}. One of the most well-studied of these is the integrable $2-$component \textsc{Camassa--Holm} system appearing in \cite{Chen2006b, Ivanov2006, Kuzmin2007},
\begin{gather}
  h_{\,t}\ +\ (h\,u)_{\,x}\ =\ 0\,, \\
  u_{\,t}\ +\ 3\,u\,u_{\,x}\ -\ u_{\,t\,x\,x}\ -\ 2\,u_{\,x}\,u_{\,x\,x}\ -\ u\,u_{\,x\,x\,x}\ +\ g\,h\,h_{\,x}\ =\ 0\,,
\end{gather}
which has been derived in the context of shallow-water theory by \textsc{Constantin} and \textsc{Ivanov} \cite{Constantin2008a} (also see \cite{Ionescu-Kruse2013a}). This system admits peakon-type solutions, and as noted in \cite{Dutykh2016}, it admits some degenerate front-type traveling wave solutions, which however necessarily have $h\ \to\ 0$ as either $x\ \to\ +\,\infty$ or $-\,\infty\,$.

The existence of weakly singular weak solutions of the rSV equations raises many interesting analytical and numerical issues that we cannot address here. For example, do smooth solutions develop weak singularities in finite time? Do finite-energy weak solutions exist globally in time? How can we approximate solutions well numerically despite weak singularities? Are weakly singular shock profiles and cusped solitary waves stable? Can similar regularization mechanisms be used to approximate shock waves in other interesting physical systems? (\emph{E.g.}, the classical \textsc{Saint-Venant} equations are formally identical to isentropic \textsc{Euler} compressible fluid equations with a particular pressure-density relation.) It would be strange if this novel and interesting phenomenon were unique to the shallow water equations. 


\section{Shock waves for the classical shallow-water system}

Let us summarize some well-known basic properties of shock-wave solutions of the classical shallow-water (\textsc{Airy} or \textsc{Saint-Venant}) system for water depth $h\,(x,\,t)\ >\ 0$ and average horizontal velocity $u\,(x,\,t)\,$:
\begin{align}
  & h_{\,t}\ +\ (h\,u)_{\,x}\ =\ 0\,,\label{e:swh}\\
  & (h\,u)_{\,t}\ +\ \paren{h\,u^{\,2}\ +\ \half\;g\,h^{\,2}}_{\,x}\ =\ 0\,.\label{e:swu}
\end{align}
This system has two \textsc{Riemann} invariants $u\ \pm\ 2\,\sqrt{\,g\,h}\,$, and two characteristic speeds 
\begin{equation*}
  \lambda_{\,1}\ =\ u\ -\ \sqrt{\,g\,h}\,, \qquad 
  \lambda_{\,2}\ =\ u\ +\ \sqrt{\,g\,h}\,.
\end{equation*}


\subsection{Jump conditions} 

A piecewise smooth solution that jumps along a curve $x\ =\ X\,(t)$ is a weak solution if and only if the \textsc{Rankine--Hugoniot} conditions hold at each point of the curve:
\begin{align}
  &-\,s\,\bracket{h}\ +\ \bracket{h\,u}\ =\ 0\,,\label{rh1}\\
  &-\,s\,\bracket{h\,u}\ +\ \bracket{h\,u^{\,2}\ +\ \half\;g\,h^{\,2}}\ =\ 0\,.\label{rh2}
\end{align}
Here $s\ =\ \dot X\,(t)$ is the jump speed and $[\,h\,]\ \eqdef\ h_{\,+}\ -\ h_{\,-}$ is the difference of right and left limits at the shock location, with similar definitions for the other brackets, \emph{e.g.}, $[\,h\,u\,]\ \eqdef\ h_{\,+}\,u_{\,+}\ -\ h_{\,-}\,u_{\,-}\,$.

After eliminating $s$ from the \textsc{Rankine--Hugoniot} conditions one finds
\begin{equation*}
  \left[\,\half\;g\,h^{\,2}\,\right]\,[\,h\,]\ =\ \frac{g\,(h_{\,+}\ +\ h_{\,-})}{2}\;[\,h\,]^{\,2}\ =\ [\,h\,u\,]^{\,2}\ -\ [\,h\,u^{\,2}\,]\,[\,h\,]\ =\ h_{\,+}\,h_{\,-}\,[\,u\,]^{\,2}\,,
\end{equation*}
so that the states $(h_{\,\pm},\,u_{\,\pm})$ lie on the \textsc{Hugoniot} curves given by
\begin{equation*}
  u_{\,+}\ -\ u_{\,-}\ =\ \pm\,\gamma\,(h_{\,+}\ -\ h_{\,-})\,, \qquad \gamma\ \eqdef\ \sqrt{\,\frac{g\,(h_{\,+}\ +\ h_{\,-})}{2\,h_{\,+}\,h_{\,-}}}\,.
\end{equation*}
Correspondingly the jump speed is determined by 
\begin{equation}\label{e:s}
  s\ =\ \frac{[\,h\,u\,]}{[\,h\,]}\ =\ u_{\,+}\ \pm\ \gamma\,h_{\,-}\ =\ u_{\,-}\ \pm\ \gamma\,h_{\,+}\,.
\end{equation}
In these relations, the $-$ sign corresponds to $1-$waves and the $+$ sign corresponds to $2-$waves. Physically meaningful shock waves satisfy the \textsc{Lax} shock conditions:
\begin{equation}\label{c:lax}
 \begin{array}{ll}
   u_{\,-}\ -\ \sqrt{\,g\,h_{\,-}}\ >\ s\ >\ u_{\,+}\ -\ \sqrt{\,g\,h_{\,+}} & \quad\mbox{for $1-$shocks,} \\ 
   u_{\,-}\ +\ \sqrt{\,g\,h_{\,-}}\ >\ s\ >\ u_{\,+}\ +\ \sqrt{\,g\,h_{\,+}} & \quad\mbox{for $2-$shocks.}
 \end{array}
\end{equation}
From \eqref{e:s} one finds that the Lax conditions hold if and only if 
\begin{equation}\label{c:lax2}
 \begin{array}{ll} 
   h_{\,-}\ <\ h_{\,+} & \mbox{for $1-$shocks,}\\
   h_{\,-}\ >\ h_{\,+} & \mbox{for $2-$shocks.}
 \end{array}
\end{equation}
The two wave families are related via the natural spatial reflection symmetry of the shallow water equations:
\begin{equation*}
  (x,\,t)\ \to\ (-\,x,\,t)\,, \qquad (h,\,u)\ \to\ (h,\,-\,u)\,.
\end{equation*}
Under this symmetry, $1-$shocks are mapped to $2-$shocks and vice versa.


\subsection{Energy dissipation}

The energy dissipation identity for a piecewise-smooth solution with shock curve $\Gamma\ =\ \{\,(x,\,t)\;:\; x\ =\ X\,(t)\,\}$ takes the form \eqref{e:eeo}, where the measure $\mu$ is absolutely continuous with respect to $1-$dimensional \textsc{Hausdorff} measure (arc length measure) restricted to the shock curve $\Gamma\,$. Denoting this \textsc{Hausdorff} measure by $\sigma\,$, in terms of the parametrization $x\ =\ X\,(t)$ we can write informally that $\ud\sigma\ =\ \sqrt{\,1\ +\ s^{\,2}}\,\ud t$ and
\begin{equation}\label{e:diss1}
  \ddd\ \eqdef\ \frac{\ud\mu}{\ud t}\ =\ -\,s\,[\,\eeo\,]\ +\ [\,\qqo\,]\ =\ \left[\half\;h\,(u\ -\ s)^{\,3}\ +\ g\,h^{\,2}\,(u\ -\ s)\right]\,.
\end{equation}
One verifies this identity by expanding $(u\ -\ s)^{\,3}$ and using that
\begin{equation*}
  s^{\,3}\,[\,h\,]\ =\ s^{\,2}\,[\,h\,u\,]\ =\ s\,\left[\,h\,u^{\,2}\ +\ \half\;g\,h^{\,2}\,\right]
\end{equation*}
from the \textsc{Rankine--Hugoniot} conditions. The precise meaning of \eqref{e:diss1} and \eqref{e:eeo} is that for any smooth test function $\varphi$ with support in a small neighborhood of the shock curve $\Gamma$ and contained in the half-plane where $x\ \in\ \R$ and $t\ >\ 0\,$, we have
\begin{equation*}
  \int_{\,0}^{\,\infty}\int_{-\,\infty}^{\,\infty} (-\,\eeo\,\D_{\,t}\,\varphi\ -\ \qqo\,\D_{\,x}\,\varphi)\,\ud\,x\,\ud\,t\ =\ \int_{\,\Gamma}\,\varphi\,\ud\mu\ =\ \int_{\,0}^{\,\infty}\,\varphi\,(X\,(t),\,t)\,\ddd\,(t)\,\ud t\,.
\end{equation*}

The identity \eqref{e:diss1} is related to the \textsc{Galilean} invariance of the shallow-water equations after changing to a frame moving with constant speed $s$ frozen at some instant of time. To conveniently compute further we introduce $v\ =\ u\ -\ s$ and write
\begin{equation}\label{d:vpm}
  v_{\,-}\ =\ u_{\,-}\ -\ s\,, \qquad v_{\,+}\ =\ u_{\,+}\ -\ s\,,
\end{equation}
and note that by the \textsc{Rankine--Hugoniot} conditions,
\begin{align}\label{d:M0}
  M\ &\eqdef\ h_{\,+}\,v_{\,+}\ =\ h_{\,-}\,v_{\,-}\,, \\ 
  \label{d:N0}
  N\ &\eqdef\ h_{\,+}\,v_{\,+}^{\,2}\ +\ \half\;g\,h_{\,+}^{\,2}\ =\ h_{\,-}\,v_{\,-}^{\,2}\ +\ \half\;g\,h_{\,-}^{\,2}\,.
\end{align}
With the same choice of sign as in \eqref{e:s} we find
\begin{align}\label{d:M}
  M\ &=\ \mp\,\gamma\,h_{\,+}\,h_{\,-}\,, \\ 
  \label{d:N}
  N\ &=\ \frac{M^{\,2}}{h_{\,\pm}}\ +\ \half\;g\,h_{\,\pm}^{\,2}\ =\ \half\;g\,(h_{\,+}^{\,2}\ +\ h_{\,+}\,h_{\,-}\ +\ h_{\,-}^{\,2})\,.
\end{align}
Then using \eqref{d:M} and \eqref{c:lax2}, we compute
\begin{equation}\label{e:dval}
  \ddd\ =\ \frac{M^{\,3}}2\;\left[\,\frac{1}{h^{\,2}}\,\right]\ +\ g\,M\,[\,h\,]\ =\ \pm\,\frac14\;g\,\gamma\,[\,h\,]^{\,3}\ <\ 0\,,
\end{equation}
for both $1-$shocks and $2-$shocks. Note that the dissipation is of the order of the amplitude cubed for small shocks.


\section{Weakly singular shock profiles for the regularized system}

Now consider any simple piecewise-constant shock-wave solution of the shallow water equations, in the form
\begin{equation}\label{d:simpleshock}
  (h,\,u)\ =\ 
  \begin{cases}
    (h_{\,-},\,u_{\,-}) & x\ <\ s\,t\,, \\ 
    (h_{\,+},\,u_{\,+}) & x\ >\ s\,t\,,
  \end{cases}
\end{equation}
where $s\,$, $h_{\,\pm}\,$, and $u_{\,\pm}$ are constants with $h_{\,\pm}\ >\ 0\,$. Our goal in this section is to show that the regularized \textsc{Saint-Venant} equations \eqref{e:rsvh} -- \eqref{e:rsvu} admit a corresponding traveling-wave solution having shock profile that is continuous and piecewise smooth, and dissipates energy at the precise rate that the corresponding classical shock does.

We mention that through the time-reversal symmetry
\begin{equation*}
  (x,\,t)\ \to\ (x,\,-t)\,, \qquad (h,\,u)\ \to\ (h,\,-u)\,,
\end{equation*}
the traveling waves that we obtain remain as valid weak solutions of the rSV system, which generate energy instead of dissipating it. These solutions correspond to non-physical shocks for the shallow-water equations that violate the \textsc{Lax} conditions in \eqref{c:lax}.


\subsection{Construction of shock profiles}

Because both the rSV and shallow water equations are invariant under spatial reflection, we may assume the shock is a $2-$shock without loss of generality. Moreover, the rSV and shallow water equations are invariant under the \textsc{Galilean} transformation taking
\begin{equation*}
  u\ \to\ u\ +\ s\,, \quad \partial_{\,t}\ \to\ -\,s\,\partial_{\,x}\ +\ \partial_{\,t}\,.
\end{equation*}
Thus it is natural to work in the frame of reference moving with the shock at speed $s$ and seek a steady wave profile that is smooth except at the origin $x\ =\ 0\,$. Adopting the notation in \eqref{d:vpm} and writing $v\ =\ u\ -\ s$ for convenience, therefore we seek time-independent functions $h\,:\ \R\ \to\ (0,\,\infty)$ and $v\,:\ \R\ \to\ \R$ such that $h$ and $v$ are continuous, smooth except at $x\ =\ 0\,$, take the limiting values
\begin{equation}\label{e:hvlim}
  (h,\,v)\ \to\ \begin{cases}
    (h_{\,-},\,v_{\,-}) & x\ \to\ -\infty\,, \\ 
    (h_{\,+},\,v_{\,+}) & x\ \to\ +\infty\,,
\end{cases}
\end{equation}
and provide a weak solution of the steady rSV equations
\begin{gather}\label{e:steadyhv}
  (h\,v)_{\,x}\ =\ 0\,, \qquad (h\,v^{\,2}\ +\ \half\;g\,h^{\,2}\ +\ \eps\,\rrr\,h^{\,2})_{\,x}\ =\ 0\,, \\
  \label{e:steadyR}
  \rrr\ =\ h\,v_{\,x}^{\,2}\ -\ h\,v\,v_{\,x\,x}\ -\ g\,(h\,h_{\,x\,x}\ +\ \half\;h_{\,x}^{\,2})\,.
\end{gather} 
As is natural, we will find solutions whose derivatives approach zero as $x\ \to\ \pm\,\infty\,$. Thus upon integration we find that 
\begin{align}
  & h\,v\ =\ M \,, \label{RH1a} \\
  & h\,v^{\,2}\ +\ \half\;g\,h^{\,2}\ +\ \eps\,\mathcal{R}\,h^{\,2}\ =\ N\,, \label{RH2a} 
\end{align}
where $M$ and $N$ are the \textsc{Rankine--Hugoniot} constants defined in \eqref{d:M0} and \eqref{d:N0} and are given by \eqref{d:M} and \eqref{d:N}, respectively.

Let us first work on the right half-line where $x\ >\ 0\,$. In terms of the dimensionless variables given by 
\begin{equation*}
  \hnon\ =\ \frac{h}{h_{\,+}}\,, \quad \vnon\ =\ \frac{v}{v_{\,+}}\,, \quad z\ =\ \frac{x}{h_{\,+}}\,,
\end{equation*}
and the squared \textsc{Froude} number on the right,
\begin{equation*}
  \fff_{\,+}\ =\ \frac{v_{\,+}^{\,2}}{g\,h_{\,+}}\,,
\end{equation*}
the equations take the form 
\begin{align*}
  & \hnon\,\vnon\ =\ 1\,, \\
  & \fff\,\hnon\,\vnon^{\,2}\ +\ \half\;\hnon^{\,2}\ +\ \eps\,\fff\,\hnon^{\,3}\,(\vnon_{\,z}^{\,2}\ -\ \vnon\,\vnon_{\,z\,z})\ -\ \eps\,(\hnon^{\,3}\,\hnon_{\,z\,z}\ +\ \half\;\hnon^{\,2}\,\hnon_{\,z}^{\,2})\ =\ \fff\ +\ \half\,.
\end{align*}
(For simplicity we temporarily drop the subscript on $\fff_{\,+}$ here.) Eliminating $\vnon$ we obtain a single equation for the dimensionless wave height $\hnon\,$,
\begin{equation*}
  \frac{\fff}{\hnon}\ +\ \half\;\hnon^{\,2}\ +\ \frac{\eps\,\fff}{\hnon}\;(\hnon\,\hnon_{\,z\,z}\ -\ \hnon_{\,z}^{\,2})\ -\ \eps\,(\hnon^{\,3}\,\hnon_{\,z\,z}\ +\ \half\;\hnon^{\,2}\,\hnon_{\,z}^{\,2})\ =\ \fff\ +\ \half\,.
\end{equation*}
Dividing this equation by $\hnon^{\,2}$ we can rewrite it as 
\begin{equation*}
  \frac{\fff}{\hnon^{\,3}}\ +\ \half\ +\ \eps\,\fff\,(\hnon\,\inv\,\hnon_{\,z})_{\,z}\,\hnon\,\inv\ -\ \eps\,(\hnon^{\,\half}\,\hnon_{\,z})_{\,z}\,\hnon^{\,\half}\ =\ \frac{\fff\ +\ \half}{\hnon^{\,2}}\,.
\end{equation*}
Further multiplying by $\hnon_{\,z}$ one can integrate this equation to obtain
\begin{equation}\label{e:odeH}
  \eps\,\hnon_{\,z}^{\,2}\ =\ G\,(\fff,\,\hnon)\ \eqdef\ \frac{(\hnon\ -\ \fff)\,(\hnon\ -\ 1)^{\,2}}{\hnon^{\,3}\ -\ \fff}\,.
\end{equation}
Here the integration constant is determined by requiring $\hnon\ \to\ 1$ as $z\ \to\ \infty\,$.

In terms of the original dimensional variables this equation takes the form
\begin{equation}\label{e:hplus}
  \eps\,h_{\,x}^{\,2}\ =\ G\,\left(\fff_{\,+},\,\frac{h}{h_{\,+}}\right)\ =\ \frac{(h\ -\ h_{\,+}\,\fff_{\,+})\,(h\ -\ h_{\,+})^{\,2}}{h^{\,3}\ -\ h_{\,+}^{\,3}\,\fff_{\,+}}\,.
\end{equation}
On the left half-line where $x\ <\ 0\,$, a similar integration procedure yields
\begin{equation}\label{e:hminus}
  \eps\,h_{\,x}^{\,2}\ =\ G\,\left(\fff_{\,-},\,\frac{h}{h_{\,-}}\right)\ =\ \frac{(h\ -\ h_{\,-}\,\fff_{\,-})\,(h\ -\ h_{\,-})^{\,2}}{h^{\,3}\ -\ h_{\,-}^{\,3}\,\fff_{\,-}}\,,
\end{equation}
with $\fff_{\,-}\ =\ v_{\,-}^{\,2}/g\,h_{\,-}\,$. We note that these equations correspond to equation (29) of \cite{Clamond2018} with the appropriate choice of integration constants.

Recalling that we are dealing with a $2-$shock for which $h_{\,+}\ <\ h_{\,-}\,$, we note
\begin{equation}\label{e:Fpm}
  h_{\,+}^{\,3}\,\fff_{\,+}\ =\ h_{\,-}^{\,3}\,\fff_{\,-}\ =\ \frac{M^{\,2}}{g}\ =\ \half\;(h_{\,+}\ +\ h_{\,-})\,h_{\,+}\,h_{\,-} \quad \in\ (h_{\,+}^{\,3},\,h_{\,-}^{\,3})\,.
\end{equation}
Therefore 
\begin{equation*}
  \fff_{\,-}\ <\ 1\ <\ \fff_{\,+}\,,
\end{equation*}
and furthermore, the denominators in \eqref{e:hplus} and \eqref{e:hminus}
vanish at the \emph{same critical height} $h_{\,c}$ satisfying $h_{\,+}\ <\ h_{\,c}\ <\ h_{\,-}\,$, where
\begin{equation}\label{d:hc}
  h_{\,c}^{\,3}\ =\ \half\;(h_{\,+}\ +\ h_{\,-})\,h_{\,+}\,h_{\,-}\ =\ \frac{M^{\,2}}g\,.
\end{equation}

On the right half line where $x\ >\ 0\,$, note the denominator in \eqref{e:hplus} changes sign from negative to positive as $h$ increases from $h_{\,+}$ past the critical height $h_{\,c}\,$, while the numerator is negative for $h_{\,+}\ <\ h\ <\ h_{\,+}\,\fff_{\,+}\,$. Because $h_{\,+}\,\fff_{\,+}\ =\ h_{\,c}^{\,3}\,/\,h_{\,+}^{\,2}\ >\ h_{\,c}\,$, this means that the right-hand side of \eqref{e:hplus} changes sign as $h$ increases past $h_{\,c}\,$: for $h$ near $h_{\,c}$ we have 
\begin{equation*}
  G\,\left(\fff_{\,+},\,\frac{h}{h_{\,+}}\right)\ >\ 0 \quad \mbox{for $h\ <\ h_{\,c}\,$}, \qquad G\,\left(\fff_{\,+},\,\frac{h}{h_{\,+}}\right)\ <\ 0 \quad \mbox{for $h\ >\ h_{\,c}$}\,.
\end{equation*}
Thus a solution of \eqref{e:hplus} taking values between $h_{\,+}$ and $h_{\,-}$ can exist only as long as $h\ <\ h_{\,c}\,$. Because we require $h\ \to\ h_{\,+}$as $x\ \to\ +\,\infty\,$, such a solution must be monotone decreasing and satsify
\begin{equation}\label{e:hp2}
  \sqrt{\eps}\,h_{\,x}\ =\ -\,\sqrt{\,G\,(\fff_{\,+},\,h/h_{\,+})}\,.
\end{equation}
Actually, we have $h\,(x)\ =\ \eta_{\,+}\,(x\,/\,\sqrt\eps)$ for a unique continuous function $\eta_{\,+}\colon[0,\,\infty)\ \to\ (0,\,\infty)$ which is a smooth decreasing solution of \eqref{e:hp2} with $\eps\ =\ 1$ for $x\ >\ 0$ and satisfies 
\begin{equation*}
  \eta_{\,+}\,(0)\ =\ h_{\,c}\,, \qquad \eta_{\,+}\,(x)\ \to\ h_{\,+} \quad\mbox{as $x\ \to\ +\,\infty\,$.}
\end{equation*}
To see that this is true, one can separate variables in \eqref{e:hp2} and determine the solution implicitly according to the relation
\begin{equation}\label{e:hpint}
  \int_{\,h}^{\,h_{\,c}}\,\frac{\ud k}{\sqrt{\,G\,(\fff_{\,+},\,k/h_{\,+})}}\ =\ \frac{x}{\sqrt\eps}\,, \quad x\ \ge\ 0\,, \quad h\ \in\ (h_{\,+},\,h_{\,c}\,]\,,
\end{equation}
since the integral converges on any interval $[\,h,\,h_{\,c}\,]\ \subset\ (h_{\,+},\,h_{\,c}\,]\,$.

On the left half line where $x\ <\ 0\,$, the reasoning is similar. The numerator in \eqref{e:hminus} is positive for $h_{\,-}\ >\ h\ >\ h_{\,-}\,\fff_{\,-}$ while the denominator changes sign from positive to negative as $h$ decreases past the critical height $h_{\,c}\,$. The solution we seek takes values between $h_{\,-}$ and $h_{\,c}\,$, satisfying 
\begin{equation}\label{e:hm2}
  \sqrt\eps\,h_{\,x}\ =\ -\,\sqrt{\,G\,(\fff_{\,-},\,h/h_{\,-})}\,.
\end{equation}
Again, we have $h\,(x)\ =\ \eta_{\,-}\,(x\,/\,\sqrt\eps)$ for a unique continuous function $\eta_{\,-}\,\colon(-\infty,\,0\,]\ \to\ (0,\,\infty)$ which is a smooth decreasing solution of \eqref{e:hm2} with $\eps\ =\ 1$ for $x\ <\ 0$ and satisfies 
\begin{equation*}
  \eta_{\,-}\,(0)\ =\ h_{\,c}\,, \qquad \eta_{\,-}\ \to\ h_{\,-} \quad\mbox{ as $x\ \to\ -\,\infty\,$.}
\end{equation*}
The solution is determined implicitly in this case according to the relation
\begin{equation}\label{e:hmint}
  \int_{\,h}^{\,h_{\,c}}\frac{\ud k}{\sqrt{\,G\,(\fff_{\,-},\,k/h_{\,-})}}\ =\ \frac{x}{\sqrt\eps}\,, \quad x\ <\ 0\,, \quad h\ \in\ (h_{\,c},\, h_{\,-})\,.
\end{equation}


\bigskip
\paragraph*{Summary.}

Let us summarize: Given the $2-$shock solution \eqref{d:simpleshock} of the shallow water equations, our corresponding weakly singular traveling wave solution of the rSV equations satisfies \eqref{e:hvlim} and takes the form 
\begin{equation}\label{e:hprofile}
  h\,(x,\,t)\ =\ \begin{cases}
   \displaystyle \eta_{\,+}\,\left(\frac{x\ -\ s\,t}{\sqrt\eps}\right) & x\ \ge\ s\,t\,, \\
   \displaystyle \eta_{\,-}\,\left(\frac{x\ -\ s\,t}{\sqrt\eps}\right) & x\ <\ s\,t\,,
  \end{cases} \qquad u\,(x,\,t)\ =\ s\ +\ \frac Mh\,,
\end{equation}
where $\eta_{\,\pm}$ are determined by $h_{\,+}$ and $h_{\,-}$ implicitly from \eqref{e:hpint} and \eqref{e:hmint} respectively with $\eps\ =\ 1\,$, using \eqref{e:Fpm} to determine $\fff_{\,\pm}\,$, and $h_{\,c}$ is given by \eqref{d:hc}.


\subsection{Behavior near the singular point and infinity}

The nature of the singularity at $x\ =\ s\,t$ for the solution above may be described as follows. For the function $G$ in \eqref{e:hplus}, because $h_{\,+}\,\fff_{\,+}\ =\ h_{\,c}^{\,3}/h_{\,+}^{\,2}$ we have
\begin{equation}\label{e:Gpinv}
  \frac{1}{G\,(\fff_{\,+},\,h/h_{\,+})}\ =\ \frac{(h^{\,3}\ -\ h_{\,c}^{\,3})\,h_{\,+}^{\,2}}{(h_{\,+}^{\,2}\,h\ -\ h_{\,c}^{\,3})\,(h\ -\ h_{\,+})^{\,2}}\ \sim\ K_{\,+}^{\,2}\,(h_{\,c}\ -\ h)
\end{equation}
as $h\ \to\ h_{\,c}\,$, where 
\begin{equation*}
  K_{\,+}^{\,2}\ =\ \frac{3\,h_{\,c}\,h_{\,+}^{\,2}}{(h_{\,c}^{\,2}\ -\ h_{\,+}^{\,2})\,(h_{\,c}\ -\ h_{\,+})^{\,2}}\,.
\end{equation*}
From this asymptotic description we infer from \eqref{e:hpint} that for small $x\ >\ 0\,$,
\begin{equation}\label{e:hp0plus}
  h_{\,c}\ -\ h\ \sim\ c_{\,+}\,x^{\,2/3}\,, \qquad h_{\,x}\ \sim\ -\,\frac{2}{3}\;c_{\,+}\,x^{\,-1/3}\,, \qquad h_{\,x\,x}\ \sim\ \frac{2}{9}\;c_{\,+}\,x^{\,-4/3}\,,
\end{equation}
where $c_{\,+}\ =\ (2\,K_{\,+}\sqrt\eps/3)^{\,-2/3}\,$.

A similar description holds on the other side of the singularity: From \eqref{e:hminus} we have
\begin{equation}\label{e:Gninv}
  \frac{1}{G\,(\fff_{\,-},\,h/h_{\,-})}\ =\ \frac{(h^{\,3}\ -\ h_{\,c}^{\,3})\,h_{\,-}^{\,2}}{(h_{\,-}^{\,2}\,h\ -\ h_{\,c}^{\,3})(h\ -\ h_{\,-})^{\,2}}\ \sim\ K_{\,-}^{\,2}\,(h\ -\ h_{\,c})
\end{equation}
as $h\ \to\ h_{\,c}\,$, where 
\begin{equation*}
  K_{\,-}^{\,2}\ =\ \frac{3\,h_{\,c}\,h_{\,-}^{\,2}}{(h_{\,-}^{\,2}\ -\ h_{\,c}^{\,2})\,(h_{\,c}\ -\ h_{\,-})^{\,2}}\,.
\end{equation*}
So for small $x\ <\ 0\,$,
\begin{equation}\label{e:hm0minus}
  h\ -\ h_{\,c}\ \sim\ c_{\,-}\,\abs{x}^{\,2/3}\,, \qquad h_{\,x}\ \sim\ -\,\frac{2}{3}\;c_{\,-}\abs{x}^{\,-1/3}\,, \qquad h_{\,x\,x}\ \sim\ -\,\frac{2}{9}\;c_{\,-}\abs{x}^{\,-4/3}\,,
\end{equation}
where $c_{\,-}\ =\ (2\,K_{\,-}\sqrt\eps/3)^{\,-2/3}\,$.

The behavior of $v$ follows by differentiation from \eqref{RH1a}. Thus we see that $h_{\,x}$ and $v_{\,x}$ are square integrable in any neighborhood of $x\ =\ 0$ (and belong to $L^{\,p}$ for $p\ <\ 3$), while $h_{\,x\,x}$ and $v_{\,x\,x}$ are not integrable functions. The singularities due to second derivatives in \eqref{RH2a} cancel however (see below), to produce the constant value $N\,$. This yields a valid distributional solution of the steady rSV equations \eqref{e:steadyhv} written in conservation form.

As $x\ \to\ \pm\infty\,$, it is straightforward to check that the limits in \eqref{e:hvlim} are achieved at an exponential rate.


\subsection{Distributional derivatives}

Because of the blow-up of $h_{\,x}$ at the origin, the distributional derivative of $h_{\,x}$ is no longer a classical function. Rather, it is a generalized function or a distribution which can be computed as follows.

We write $h_{\,x\,x}$ to denote the distributional derivative of $h_{\,x}$ and write $\overline{h_{\,x\,x}}$ for the classical derivative of $h_{\,x}$ that is not defined at $0\,$. Let $\varphi\ \in\ C_{\,c}^{\,\infty}\,(\RR)$ be a test function with support $\supp\varphi\ \subset\ (-L,\,L)\,$. Let $\tau$ be a subtracting operator acting on functions from $\RR$ to $\RR$ such that
\begin{equation*}
  \tau\,\varphi\,(x)\ =\ \varphi\,(x)\ -\ \varphi\,(0)\,.
\end{equation*}
Then the distributional pairing of $\varphi$ with the distribution $h_{\,x\,x}$ is 
\begin{eqnarray}
  \angl{h_{\,x\,x},\,\varphi}\ &=&\ -\int_\RR\,h_{\,x}\,\varphi_{\,x}\,\ud x\ =\ -\,\int_\RR\,h_{\,x}\,(\tau\,\varphi)_{\,x}\,\ud x \nonumber\\
  &=&\ -\lim_{\eps\ \to\ 0^{\,+}}\,\paren{\int_{\,-L}^{\,-\eps}\,h_{\,x}\,(\tau\,\varphi)_{\,x}\,\ud x\ +\ \int_{\,\eps}^{\,L}\,h_{\,x}\,(\tau\,\varphi)_{\,x}\,\ud x}\nonumber\\
  &=&\ -\,h_{\,x}\,(-L)\,\varphi\,(0)\ +\ h_{\,x}\,(L)\,\varphi\,(0)\ +\ \int_{\,-L}^{\,L}\,\overline{h_{\,x\,x}}\,(\tau\,\varphi)\,\ud x\,,
\end{eqnarray}
where in the last step we use the fact that $(\tau\,\varphi)\,(x)\ \sim\ x\,\varphi_{\,x}\,(0)$ when $x$ is small and the fact that $\overline{h_{\,x\,x}}\,\tau\,\varphi$ is integrable near $0\,$. Furthermore, the above equality is true
for all $L$ large enough, so sending $L$ to infinity we have that
\begin{equation*}
  \angl{h_{\,x\,x},\,\varphi}\ =\ \int_\RR\,\overline{h_{\,x\,x}}\,(\tau\,\varphi)\,\ud x\,.
\end{equation*}

Due to this result, the distribution $h_{\,x\,x}\,(h\ -\ h_{\,c})$ satisfies
\begin{eqnarray*}
  \angl{h_{\,x\,x}\,(h\ -\ h_{\,c}),\,\varphi}\ &=&\ \angl{h_{\,x\,x},\,(h\ -\ h_{\,c})\,\varphi} \nonumber \\
  &=&\ \int_\RR\,\overline{h_{\,x\,x}}\,\tau\,((h\ -\ h_{\,c})\,\varphi)\,\ud x\ =\ \int_\RR\,\overline{h_{\,x\,x}}\,(h\ -\ h_{\,c})\,\varphi\,\ud x \\
  &=&\ \angl{\overline{h_{\,x\,x}}\,(h\ -\ h_{\,c}),\,\varphi}
\end{eqnarray*}
where the first line is justified by the fact that $h_{\,x\,x}$ is a continuous linear functional on $W^{\,1,\,p}\,(\RR)$ for any $p\ \in\ (1,\,\infty)\,$. This implies that in the sense of distributions,
\begin{equation}\label{e:hxxbar}
  h_{\,x\,x}\,(h\ -\ h_{\,c})\ =\ \overline{h_{\,x\,x}}\,(h\ -\ h_{\,c})\,,
\end{equation}
where the right-hand side is a locally integrable function.

From this we can find a locally integrable representation of the quantity $h^{\,2}\,\rrr$ from \eqref{e:steadyR}. Differentiating \eqref{RH1a} twice and multiplying by $h^{\,2}\,v\,$, we find $h^{\,3}\,v_{\,x}^{\,2}\ =\ M^{\,2}\,h_{\,x}^{\,2}/h$ and 
\begin{equation*}
  -\,h^{\,3}\,v\,v_{\,xvx}\ =\ M^{\,2}\,\left(h_{\,x\,x}\ -\ \frac{2\,h_{\,x}^{\,2}}\;h\right)\,.
\end{equation*}
Because $M^{\,2}\ =\ g\,h_{\,c}^{\,3}\,$, using \eqref{e:hxxbar} it follows
\begin{equation*}
  h^{\,2}\,\rrr\ =\ g\,(h_{\,c}^{\,3}\ -\ h^{\,3})\,\overline{h_{\,x\,x}}\ -\ \frac{g}{2\,h}\;(2\,h_{\,c}^{\,3}\ +\ h^{\,3})\,h_{\,x}^{\,2}\,.
\end{equation*}
So we conclude that the singularities appearing in $h_{\,x\,x}$ and $v_{\,x\,x}$ do cancel each other in a way that makes the stationary momentum flux locally integrable with distributional derivative $0\,$.

Another way to see this cancellation is that the singular terms in $h^{\,2}\,\rrr$ sum up to give
\begin{eqnarray*}
  h^{\,3}\,v\,v_{\,x\,x}\ +\ g\,h^{\,3}\,h_{\,x\,x}\ &=&\ \paren{h^{\,3}\,v\,v_{\,x}\ +\ g\,h^{\,3}\,h_{\,x}}_{\,x}\ -\ (h^{\,3}\,v)_{\,x}\,v_{\,x}\ -\ g\,(h^{\,3})_{\,x}\,h_{\,x} \\
  &=&\ \paren{h^{\,3}\,v\,\paren{-\frac{M}{h^{\,2}}\,h_{\,x}}\ +\ g\,h^{\,3}\,h_{\,x}}_{\,x}\ -\ (h^{\,3}\,v)_{\,x}\,v_{\,x}\ -\ g\,(h^{\,3})_{\,x}\,h_{\,x} \\
  &=&\ g\,\paren{(h^{\,3}\ -\ h_{\,c}^{\,3})\,h_{\,x}}_{\,x}\ -\ (h^{\,3}\,v)_{\,x}\,v_{\,x}\ -\ g\,(h^{\,3})_{\,x}\,h_{\,x}\,,
\end{eqnarray*}
in which every term is a locally integrable function.


\subsection{Energy dissipation of weakly singular waves}

Here our aim is to show that the regularized shock-wave solutions of the rSV equations that correspond to the simple shallow-water shock \eqref{d:simpleshock} satisfy the distributional identity 
\begin{equation}\label{e:disseps}
  \eep_{\,t}\ +\ \qqp_{\,x}\ =\ \mu\,,
\end{equation}
where the dissipation measure $\mu$ is a constant multiple of $1-$dimensional \textsc{Hausdorff} measure restricted to the simple shock curve $\{(x,\,t)\,:\ x\ =\ s\,t\}\,$, satisfying 
\begin{equation*}
  \ddd\ =\ \frac{\ud \mu}{\ud t}\ =\ \pm\,\frac14\;g\,\gamma\,(h_{\,+}\ -\ h_{\,-})^{\,3}\ <\ 0\,,
\end{equation*}
\emph{exactly the same as the simple shallow-water shock in \eqref{d:simpleshock}}.

Indeed, the steady solution constructed above is a smooth solution of the rSV equations \eqref{e:rsvh} -- \eqref{e:rsvu} on both the right and left half-lines, hence satisfies the conservation law \eqref{e:rsvE} except at 
$x\ =\ 0\,$. In this time-independent situation this means 
\begin{equation*}
  \qqp_{\,x}\ =\ 0\,, \qquad x\ \in\ \R\ \setminus\ \{\,0\,\}\,.
\end{equation*}
Now, integration of this equation separately on the right and left half lines yields
\begin{equation*}
  \qqp\ =\ \begin{cases}
    \qqq_{\,-}\,, & x\ <\ 0\,, \\
    \qqq_{\,+}\,, & x\ >\ 0\,,
  \end{cases}
\end{equation*}
where the constants $\qqq_{\,\pm}$ can be evaluated by taking $x\ \to\ \pm\,\infty$ in the expression for $\qqp$ in \eqref{e:rsvE} and invoking the limits in \eqref{e:hvlim}. The result is that the constants $\qqq_{\,\pm}$ take the same values as appear in \eqref{e:diss1} for the simple shallow-water shock. Namely,
\begin{equation*}
  \qqq_{\,\pm}\ =\ \frac12\;h_{\,\pm}\,v_{\,\pm}^{\,3}\ +\ g\,h_{\,\pm}^{\,2}\,v_{\,\pm}\,.
\end{equation*}
Therefore, by the same calculation that leads to \eqref{e:dval}, the weak derivative of $\qqp$ on all of $\R$ is a multiple of the \textsc{Dirac} delta measure $\delta_{\,0}$ at $x\ =\ 0\,$, satisfying
\begin{equation*}
  \qqp_{\,x}\ =\ (\qqq_{\,+}\ -\ \qqq_{\,-})\,\delta_{\,0}\ =\ \ddd\,\delta_{\,0}\,,
\end{equation*}
where $\ddd$ is the same as in \eqref{e:dval}. By undoing the Galilean transformation to the frame moving with the simple shock speed, we obtain \eqref{e:disseps} with dissipation measure $\mu$ exactly as claimed above.


\section{Cusped solitary waves for the regularized system}

The construction of weakly singular shock profiles in the previous section also enables us to describe cusped solitary waves for the rSV equations. These are weak traveling-wave solutions whose limits as $x\ \to\ -\,\infty$ are the same as those as $x\ \to\ +\,\infty\,$.

The point is that weak solutions of the steady rSV equations \eqref{e:steadyhv}--\eqref{e:steadyR} can be constructed \emph{by reflection} from either piece $\eta_{\,\pm}$ of the $2-$shock profile in the previous section. For each of these pieces, the quantities on the left-hand sides in \eqref{RH1a} and \eqref{RH2a} are locally integrable (in total, though not term-wise) and indeed constant on $\R\ \setminus\ \{\,0\,\}\,$. Thus the construction above yields two valid distributional solutions of the steady rSV equations with height profiles
\begin{equation}\label{e:solwpm}
  h\,(x,\,t)\ =\ \eta_{\,\pm}\,\left(\frac{\abs{x\ -\ s\,t}}{\sqrt\eps}\right)\,,
\end{equation}
respectively satisfying $h\,(x,\,t)\ \to\ h_{\,\pm}$ as $\abs{x}\ \to\ \infty\,$. The energy of these solitary wave solutions satisfies the conservation law \eqref{e:rsvE} without alteration.


\subsection{Solitary waves of elevation}

We note that for the solution using $\eta_{\,+}\,$, the value of $h_{\,-}$ has no direct interpretation in terms of the wave shape. However, from \eqref{e:Gpinv} we see that the solitary-wave height profile with the $+$ sign can be determined from any independently chosen values of $h_{\,\infty}\ \eqdef\ h_{\,+}$ and $h_{\,c}$ with 
\begin{equation*}
  0\ <\ h_{\,\infty}\ <\ h_{\,c}\,.
\end{equation*}
Here $h_{\,c}$ is the maximum height of the wave and $h_{\,\infty}$ is the limiting value at $\infty\,$. The wave everywhere is a wave of elevation, with $h_{\,\infty}\ <\ h\,(x,\,t)\ \le\ h_{\,c}\,$, determined implicitly as in \eqref{e:hpint} and \eqref{e:Gpinv} by 
\begin{equation}\label{e:solpint}
  \int_{\,h}^{\,h_{\,c}}\,\left({\frac{h_{\,c}^{\,3}\ -\ k^{\,3}}{h_{\,c}^{\,3}\ -\ h_{\,\infty}^{\,2}\,k}}\right)^{\,\half}\;\frac{h_{\,\infty}}{k\ -\ h_{\,\infty}}\;\ud k\ =\ \frac{\abs{x\ -\ s\,t}}{\sqrt\eps}\,, \qquad x\ \in\ \R\,, \quad h\ \in\ (h_{\,\infty},\,h_{\,c}\,]\,.
\end{equation}
It is natural for solitary waves to consider $u_{\,+}\ =\ 0$ to be the limiting velocity as $\abs{x}\ \to\ \infty$ in the original frame. Then by \eqref{e:s}, $v_{\,+}\ =\ -\,s\ =\ -\gamma\,h_{\,-}\,$, whence we find using \eqref{d:hc} that 
$\gamma\ =\ \sqrt{g\,h_{\,c}}\,h_{\,c}/(h_{\,+}\,h_{\,-})$ and 
\begin{equation}\label{e:sols}
  s\ =\ \sqrt{g\,h_{\,c}}\,\frac{h_{\,c}}{h_{\,\infty}}\,.
\end{equation}
This determines the velocity profile according to 
\begin{equation}\label{e:solu}
  u\,(x,\,t)\ =\ s\ +\ \frac{M}{h}\ =\ s\,\left(1\ -\ \frac{h_{\,\infty}}{h}\right)\,.
\end{equation}
This velocity is everywhere positive, as a consequence of the fact that we started with a $2-$shock profile. We note that these solitary waves travel to the right, with speed $s$ that exceeds the characteristic speed $\sqrt{g\,h_{\,\infty}}$ at the constant state $(h_{\,\infty},\,0)$ in this case. The spatial reflection symmetry yields solitary waves that travel to the left instead. This symmetry also recovers the solitary waves that can be constructed from $1-$shock profiles.


\subsection{Solitary waves of depression}

We obtain solitary waves of depression by using $\eta_{\,-}$ in \eqref{e:solwpm} instead of $\eta_{\,+}\,$, choosing $h_{\,\infty}\ \eqdef\ h_{\,-}$ (the wave height at $\infty$) and $h_{\,c}$ (the minimum wave height) arbitrary subject to the requirement that
\begin{equation*}
  0\ <\ h_{\,c}\ <\ h_{\,\infty}\,.
\end{equation*}
Similarly to \eqref{e:solpint}, the wave height $h\,(x,\,t)\ \in\ [h_{\,c},\,h_{\,\infty})$ is determined implicitly by 
\begin{equation}\label{e:solmint}
  \int_{\,h_{\,c}}^{\,h}\,\left({\frac{k^{\,3}\ -\ h_{\,c}^{\,3}}{h_{\,\infty}^{\,2}\,k\ -\ h_{\,c}^{\,3}}}\right)^{\,\half}\;\frac{h_{\,\infty}}{h_{\,\infty}\ -\ k}\,\ud k\ =\ \frac{\abs{x\ -\ s\,t}}{\sqrt\eps}\,, \qquad x\ \in\ \R\,, \quad h\ \in\ [\,h_{\,c},\,h_{\,\infty})\,.
\end{equation}
Considering $u_{\,-}\ =\ 0$ to be the limiting velocity as $\abs{x}\ \to\ \infty\,$, we find $v_{\,-}\ =\ -\,s\ =\ -\,\gamma\,h_{\,+}$ from \eqref{e:s}, whence
the solitary wave speed is again given by equation \eqref{e:sols}, and again the corresponding velocity profile is given by \eqref{e:solu}. This time, the velocity is everywhere negative (when starting with the $2-$shock profile), while the solitary wave travels to the right ($s\ >\ 0$) but with speed $s$ \emph{less than} the characteristic speed $\sqrt{g\,h_{\,\infty}}$ of the state at infinity. Again, spatial reflection yields waves of depression traveling to the left.


\section{Parametric formulae for shock profiles and cusped waves}

Here we describe how weakly singular shock profiles and cusped waves can be determined in a parametric form,
\begin{equation*}
  h\ =\ h\,(\xi)\,, \qquad x\ =\ x\,(\xi)\,, \qquad \xi\ \in\ \R\,,
\end{equation*}
by a quadrature procedure that eliminates having to deal with the singularities present in the ODEs \eqref{e:hplus}, \eqref{e:hminus} and in the integrands of the implicit relations \eqref{e:hpint}, \eqref{e:hmint}, \eqref{e:solpint}. Inspired by the fact that classical solitary wave profiles of the form $f\,(\xi)\ =\ \beta\,\sech^{\,2}\,(\frac12\;\xi)$ (and their translates) satisfy an equation with cubic polynomial as right-hand side, 
\begin{equation}\label{e:sech2}
  f_{\,\xi}^{\,2}\ =\ f^{\,2}\,\left(1\ -\ \frac{f}{\beta}\right)\,,
\end{equation}
we modify the dimensionless ODE~\eqref{e:odeH} by replacing $H^{\,3}$ in the denominator by its asymptotic value $1\,$. Thus we seek the solution of \eqref{e:odeH} in parametric form $H\ =\ H\,(\xi)\,$, $z\ =\ z\,(\xi)$ by solving 
\begin{gather}\label{e:Hpar}
  H_{\,\xi}^{\,2}\ =\ \frac{(\hnon\ -\ \fff)\,(\hnon\ -\ 1)^{\,2}}{1\ -\ \fff}\ =\ (H\ -\ 1)^{\,2}\,\left(1\ -\ \frac{H\ -\ 1}{\fff\ -\ 1}\right)\,, \\
  \label{e:zpar}
  z_{\,\xi}^{\,2}\ =\ \eps\,\frac{H^{\,3}\ -\ \fff}{1\ -\ \fff}\,.
\end{gather}

We require $H^{\,3}\ =\ \fff$ when $z\ =\ 0\,$. It is convenient to require $z\,(0)\ =\ 0\,$. Comparing the form of \eqref{e:Hpar} with \eqref{e:sech2} we find the appropriate solution of \eqref{e:Hpar} on either half-line $\xi\ \ge\ 0$ or $\xi\ \le\ 0$ can be written in the form 
\begin{equation}\label{e:Hsol}
  H\,(\xi)\ =\ 1\ +\ (\fff\ -\ 1)\,\sech^{\,2}\,\left(\frac12\;\abs{\xi}\ +\ \alpha\right)\,,
\end{equation}
where $H\,(0)\ =\ \fff^{\,1/3}$ provided 
\begin{equation*}
  \cosh^{\,2}\,\alpha\ =\ \frac{\fff\ -\ 1}{\fff^{\,1/3}\ -\ 1}\,.
\end{equation*}
A unique $\alpha\ >\ 0$ solving this equation exists in either case $\fff\ >\ 1$ or $0\ <\ \fff\ <\ 1\,$, namely
\begin{equation}\label{d:alpha}
  \alpha\ =\ \ln\,(\sqrt{\gamma}\ +\ \sqrt{\gamma\ -\ 1})\,, \qquad \gamma\ =\ \frac{\fff\ -\ 1}{\fff^{\,1/3}\ -\ 1}\,,
\end{equation}
because $\gamma\ >\ 1\,$. Now $z\,(\xi)$ is recovered by quadrature from \eqref{e:zpar} as
\begin{equation}\label{e:zsol}
  z\,(\xi)\ =\ \sqrt\eps\;\int_{\,0}^{\,\xi}\,\left(\frac{H\,(\zeta)^{\,3}\ -\ \fff}{1\ -\ \fff}\right)^{\,1/2}\,\ud\zeta\,.
\end{equation}
To express this result in dimensional terms for $h\ =\ h_{\,\pm}\,H$ in each case as appropriate, we recall $\fff_{\,\pm}\ =\ h_{\,c}^{\,3}/h_{\,\pm}^{\,3}$ where $h_{\,c}$ may be determined from $h_{\,+}\,$, $h_{\,-}$ by \eqref{d:hc}. We obtain
\begin{gather}\label{e:hint}
  h\,(\xi)\ =\ h_{\,\pm}\ +\ \frac{(h_{\,c}\ -\ h_{\,\pm})\,\cosh^{\,2}\,\alpha_{\,\pm}}{\cosh^{\,2}\,(\frac12\;\abs{\xi}\ +\ \alpha_{\,\pm})}\,, \\
  \label{e:xint}
  x\,(\xi)\ =\ \sqrt\eps\,h_{\,\pm}\,\int_{\,0}^{\,\xi}\left(\frac{h\,(\zeta)^{\,3}\ -\ h_{\,c}^{\,3}}{h_{\,\pm}^{\,3}\ -\ h_{\,c}^{\,3}}\right)^{\,1/2}\,\ud\zeta\,,
\end{gather}
where $\alpha_{\,\pm}$ is determined from \eqref{d:alpha} using $\fff\ =\ \fff_{\,\pm}\,$.

Cusped solitary waves profiles are expressed parametrically by the same formulae
after replacing $h_{\,\pm}$ with $h_{\,\infty}\,$.

An explicit expression for $x\,(\xi)$ remains to be obtained. Even if this expression could be obtained in closed form, it likely would involve special functions that may not be easily computed. In any case, it is straightforward to compute $x\,(\xi)$ directly from the integral by an efficient quadrature method. We note, however, that \textsc{Taylor} expansion of $\sech^{\,2}\,(\frac12\;\abs{\xi}\ +\ \alpha_{\,\pm})$ implies that for small $\abs{\xi}\,$,
\begin{equation*}
  \frac{h\,(\xi)\ -\ h_{\,c}}{h_{\,\pm}\ -\ h_{\,c}}\ =\ \abs{\xi}\,\tanh\,\alpha_{\,\pm}\ +\ \O\,(\abs{\xi}^{\,2})\,.
\end{equation*}
Consequently the integrand of \eqref{e:xint} has a weak singularity at $0\,$, with
\begin{equation*}
  \left(\frac{h\,(\zeta)^{\,3}\ -\ h_{\,c}^{\,3}}{h_{\,\pm}^{\,3}\ -\ h_{\,c}^{\,3}}\right)^{\,1/2}\ =\ K\,\abs{\zeta}^{\,1/2}\ +\ \O\,(\abs{\zeta})\,, \qquad K\ =\ \left(\frac{3\,h_{\,c}^{\,2}\,\tanh\,\alpha_{\,\pm}}{h_{\,\pm}^{\,2}\ +\ h_{\,\pm}\,h_{\,c}\ +\ h_{\,c}^{\,2}}\right)^{\,1/2}\,.
\end{equation*}
This singularity can be eliminated by a change of variable $\zeta\ =\ \pm\,y^{\,2}$ --- then simple quadratures will yield accurate numerical approximations.


\section{Numerical simulations}
\label{sec:6}

In this section we examine how the theory of weakly singular shock profiles
developed in this paper fits the smoothed shocks observed 
in the computations carried out in \cite{Clamond2018}.


\subsection{A dynamically generated wave front}

\begin{figure}
\begin{center}
\includegraphics[width=0.95\linewidth]{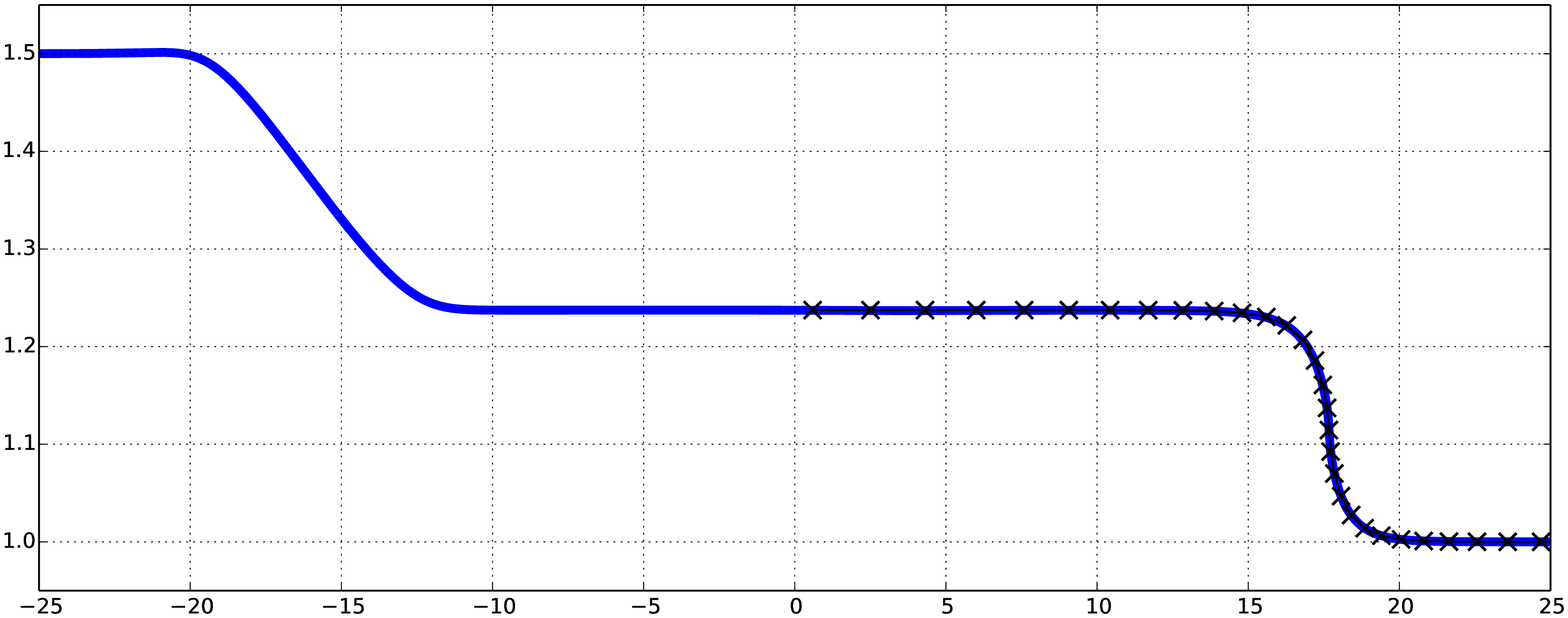}
\put(-340,90){\large $h$}
\put(-50,0){\large $x$}

\includegraphics[width=0.95\linewidth]{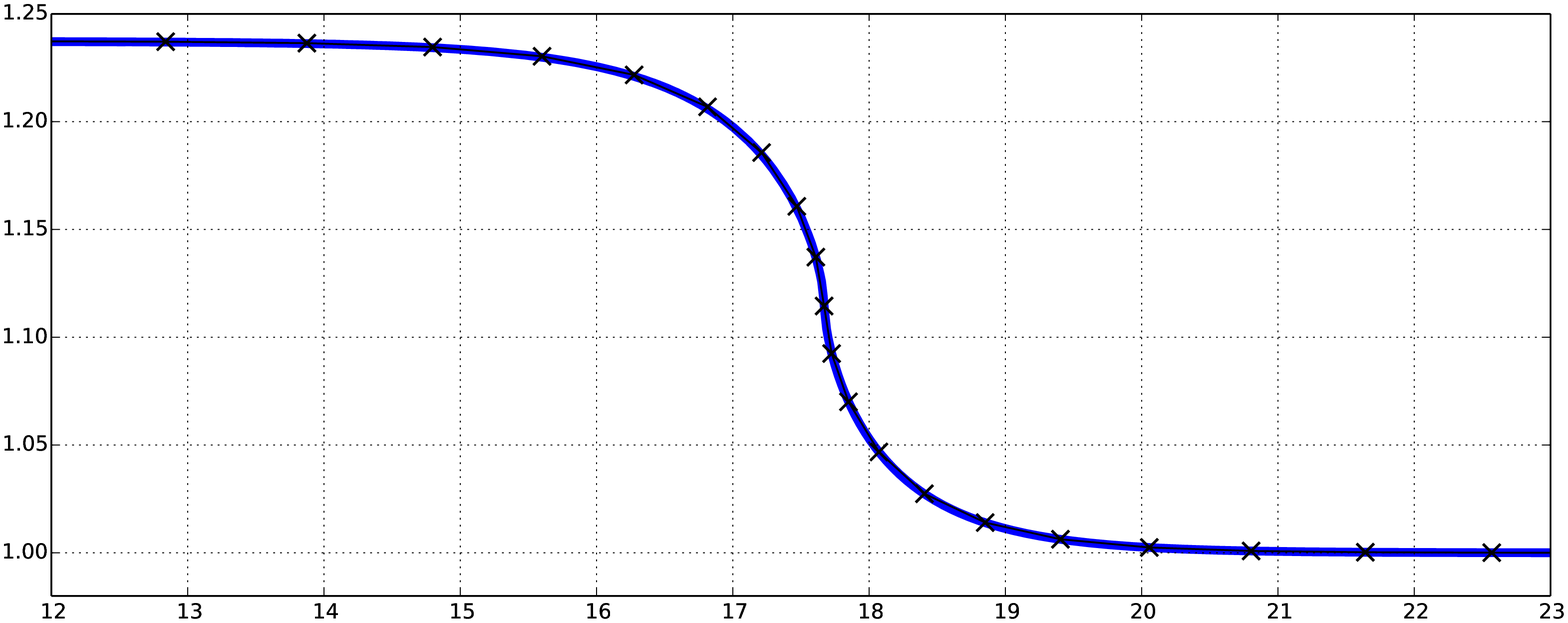}
\put(-340,100){\large $h$}
\put(-50,0){\large $x$}

\caption{\small\em Comparison of shock profile with dam-break computation of \cite{Clamond2018}. The solid line is the rSV solution with $\eps\ =\ 0.5$ computed by a pseudospectral method. Crosses mark the shock profile computed as in \eqref{e:hint} -- \eqref{e:xint}, shifted by $17.67\,$.}
\label{fig:1} 
\end{center}
\end{figure}

In Fig.~\ref{fig:1} we compare a shock profile computed by the theory developed in this paper with a solution to the rSV system computed as in \cite{Clamond2018} for ``dam-break'' initial data, similar to a \textsc{Riemann} problem for the shallow-water system. For a recent treatment of the classical Riemann problem for the shallow-water equations, including a discussion of analytical properties as well as numerical techniques, see \cite{Holden2015}.

The solid line in Fig.~\ref{fig:1} is from the numerically computed solution to the rSV system at time $t\ =\ 15$ with $\eps\ =\ 0.5$ and smoothed step function (``dam break'') initial data
\begin{equation*}
  h_{\,0}\,(x)\ =\ h_{\,-}\ +\ \frac12\;(h_{\,+}\ -\ h_{\,-})\,(1\ +\ \tanh(\delta\,x))
\end{equation*}
for $h_{\,-}\ =\ 1.5\,$, $h_{\,+}\ =\ 1\,$, $g\ =\ 1\,$, $\delta\ =\ 1\,$, as indicated in \cite{Clamond2018}. The numerical computation was performed with a \textsc{Fourier} pseudospectral method as described in \cite{Dutykh2011a}, using an Erfc-Log filter for anti-aliasing \cite{Boyd1995} and with $N\ =\ 8192$ modes on a periodic domain of length $4\,L$ with $L\ =\ 25\,$.

The crosses mark the shock profile solution computed parametrically using formulae \eqref{e:hint} -- \eqref{e:xint} of the previous section with $h_{\,-}\ =\ 1.2374$ and $h_{\,+}\ =\ 1\,$, with $x$ shifted by $17.67\,$. The bottom part of the figure is a zoom-in on the indicated region of the upper part. We remark that the computed rSV solution in Fig.~\ref{fig:1} corresponds directly to Fig.~3(c) of \cite{Clamond2018} --- due to a late change of notation the values of $\eps$ reported for the computations in \cite{Clamond2018} correspond to $2\,\eps$ in the present notation.

\subsection{Energy dissipation}

\begin{figure}
\begin{center}
  \includegraphics[width=0.95\linewidth]{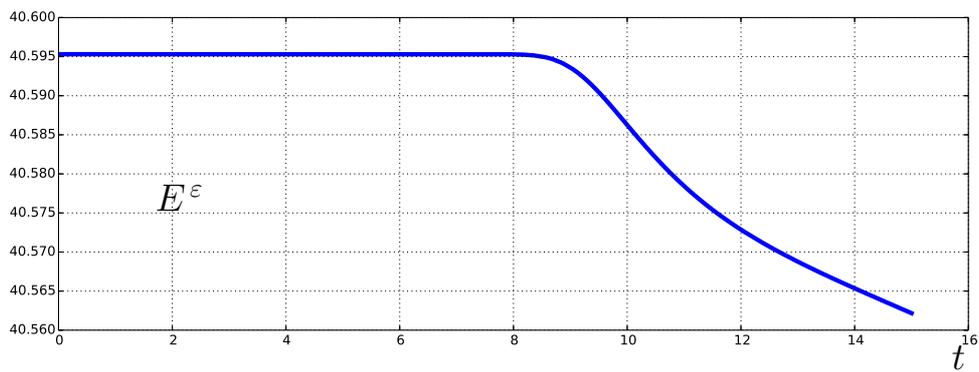}
  \put(-348,60){\large $E^{\,\eps}$}
  \put(-50,0){\large $t$}
  \caption{\small\em Total energy $E^{\,\eps}$ vs. $t$ in the smoothed dam break problem as in Fig.~\ref{fig:1} with $\eps\ =\ 0.5\,$.}
  \label{fig:2}
\end{center}
\end{figure}

In Fig.~\ref{fig:2} we plot the total energy from \eqref{e:rsvE},
\begin{equation}\label{d:totalE}
  E^{\,\eps}\,(t)\ =\ \int_{\,-L}^{\,L} \eep\,\ud x\,,
\end{equation}
as a function of time, for a solution computed as in Fig.~\ref{fig:1} but with anti-aliasing performed using the filter employed by \textsc{Hou} and \textsc{Li} in \cite{Hou2007}, namely 
\begin{equation*}
  \rho\,(2\,k\,/\,N)\ =\ \exp(\,-36\,\abs{2\,k\,/\,N}^{\,36})\,, \quad k\ =\ -N/2,\,\ldots,\,N/2\ -\ 1\,,
\end{equation*}
applied on each time step. From this data, we estimate the average energy decay rate $\ud E^{\,\eps}/\ud t\ \approx\ -\,0.00326$ over the range $t\ \in\ [\,14,\,15\,]\,$. Corresponding to $h_{\,-}\ =\ 1.2374\,$, $h_{\,+}\ =\ 1\,$, the dissipation formula \eqref{e:dval} predicts $\ud E^{\,\eps}/\ud t\ =\ -\,0.00318\,$, giving a relative error of less than $2.6$ percent.

\subsection{Cusped waves}

\begin{figure}
\begin{center}
  \includegraphics[width=0.9\linewidth]{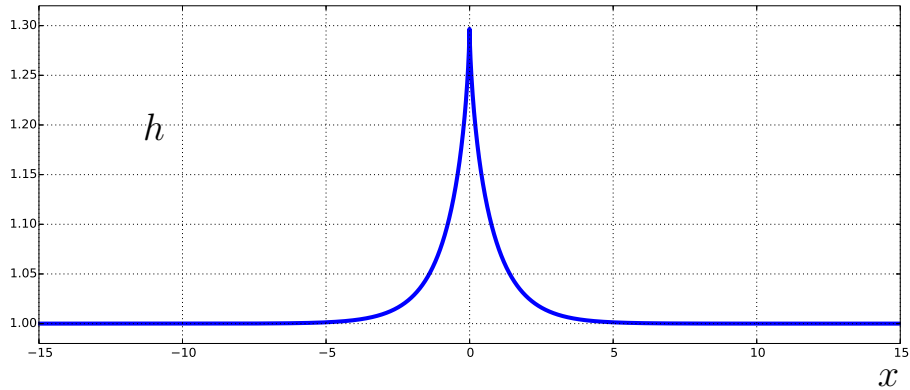}
  \put(-325,93){\large $h$}
  \put(-50,0){\large $x$}
  \caption{\small\em Cusped solitary wave profile for $h_{\,\infty}\ = 1\,$, $h_{\,c}\ =\ 1.3\,$.}
  \label{fig:cusp}
\end{center}
\end{figure}

\begin{figure}
  \begin{center}
  \includegraphics[width=0.9\linewidth]{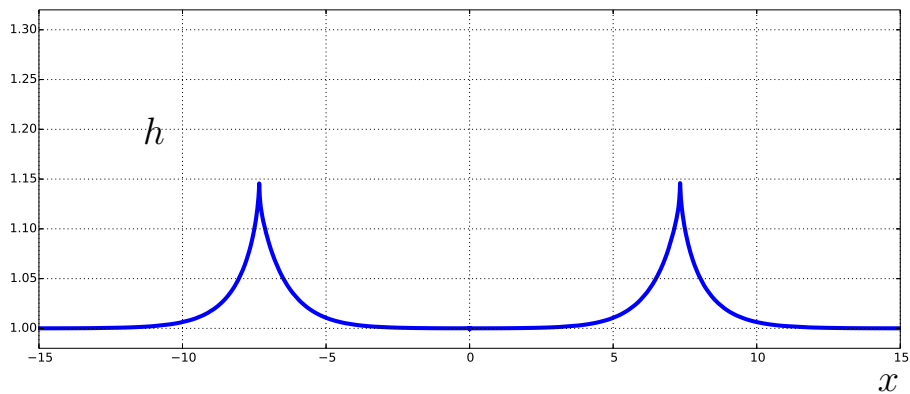}
  \put(-325,93){\large $h$}
  \put(-50,0){\large $x$}
  \caption{\small\em Numerical solution at $t\ =\ 6$ with initial height from Fig.~\ref{fig:cusp}, initial velocity zero.}
  \label{fig:2peak}
  \end{center}
\end{figure}

The profile of a cusped solitary wave of elevation is plotted in Fig.~\ref{fig:cusp} for $h_{\,\infty}\ =\ h_{\,+}\ =\ 1$ and maximum height $h_{\,c}\ =\ 1.3\,$. We were not able to compute a clean isolated traveling cusped wave by taking the numerically computed wave profile for $(h,\,u)$ as initial data on a regular grid. Indeed, there is no particular reason our pseudospectral code should work well for such a singular solution, and anyway it may not be numerically stable. However, when taking the $h-$profile in Fig.~\ref{fig:cusp} as initial data with \emph{zero} initial velocity, the numerical solution develops two peaked waves traveling in opposite direction as indicated Fig.~\ref{fig:2peak}. While hardly conclusive, this evidence suggests that cusped solutions may be relevant in the dynamics of the rSV system.

The two peaks here are slightly skewed compared to the profile of a cusped solitary wave. Our limited exploration uncovered no convincing evidence that cusped waves collide ``cleanly'' enough to justify calling them `cuspons' or suggest that the rSV system is formally integrable --- It may be difficult to tell, though, as perturbed cusped waves do not leave behind a dispersive ``tail'' in this non-dispersive system.

\section{Discussion and outlook}

Our analysis of traveling wave profiles for the rSV system proves that, as the authors of \cite{Clamond2018} stated, the regularized system admits `smoothed shocks' that propagate at exactly the same speed as corresponding classical discontinuous shocks for the shallow water equations. The new waves are indeed piecewise smooth and continuous, but have weak singularities which correctly generate the same energy dissipation as the classical shocks.

This ability of the rSV system to correctly model shock wave propagation non-dispersively without oscillations while conserving energy for smooth solutions is an interesting feature which deserves further investigation. As demonstrated in \cite{Clamond2018}, it means that a rather straightforward pseudospectral method (albeit one which involves careful dealiasing, and iteration to eliminate the time derivative term in $\calR$) computes shock speeds accurately over a wide range of values of $\eps\,$, with $2\,\eps$ ranging from $0.001$ to $5$ in the examples treated in \cite{Clamond2018}.

The comparisons made in the previous section above make it plausible that the pseudospectral method used to produce Figs.~\ref{fig:1} and \ref{fig:2} is computing an accurate approximation to a solution of the rSV system which ceases to conserve energy (hence loses smoothness) around $t\ =\ 7$ or $8\,$, and develops afterward a traveling wave whose shape closely matches a weakly singular shock profile. We speculate that an important source of energy dissipation in this pseudospectral computation may be the damping of high frequency components induced for dealiasing purposes.

How this actually happens and what it may mean with regard to the design and accuracy of numerical approximations remains to be investigated in detail. Often, energy conservation, or preservation of some variational (\textsc{Lagrangian}) or symplectic (\textsc{Hamiltonian}) structure, is a desirable feature of a numerical scheme designed for long-time computations in an energy-conserving system. (See \cite{Marsden1998, Lew2004, Hairer2002, Clamond2007} for discussion of variational and symplectic integrators.) But for the rSV system considered here, exact conservation of energy appears to be {\it inappropriate} for approximating solutions containing weakly singular shock profiles, which dissipate energy as we have shown.

At present, the issue of preservation of symplectic structure may be moot anyways, since we are not aware of a canonical Hamiltonian structure for the rSV system. It seems worth mentioning, however, that the rSV system admits the following non-canonical \textsc{Hamiltonian} structure. Namely, with
\begin{equation*}
  \calH\ =\ \int\half\;h\,u^{\,2}\ +\ \half\;g\,(h\ -\ h_{\,\infty})^{\,2}\ +\ \eps\,\left(\half\;h^{\,3}\,u_{\,x}^{\,2}\ +\ \half\;g\,h^{\,2}\,h_{\,x}^{\,2}\right)\,\ud x\,,
\end{equation*}
and $m\ =\ h\,u\ -\ \eps\,(h^{\,3}\,u_{\,x})_{\,x}\,$, the rSV system is formally equivalent to
\begin{equation*}
  \partial_{\,t}\,\begin{pmatrix} m \\ h \end{pmatrix}\ =\ 
  -\,\begin{pmatrix}
    \partial_{\,x}\,m\ +\ m\,\partial_{\,x} & h\,\partial_{\,x} \\
    \partial_{\,x} h & 0
  \end{pmatrix}\cdot
  \begin{pmatrix} 
    \delta \calH/\delta m \\
    \delta \calH/\delta h
  \end{pmatrix}\,.
\end{equation*}
This is a simple variant of the \textsc{Hamiltonian} structure well-known for the \textsc{Green}--\textsc{Naghdi} equations \cite{Holm1988, Constantin1997, Johnson2002, Li2002}, obtained by replacing the \textsc{Green}--\textsc{Naghdi} \textsc{Hamiltonian} with a \textsc{Hamiltonian} derived from \eqref{d:eep}.

Finally, as we have mentioned, quite a number of analytic questions remain for further investigation, involving the development of weak singularities in smooth solutions of the rSV system, the existence of solutions with weak singularities, and whether these phenomena occur for other important models of physical systems.


\subsection*{Acknowledgments}
\addcontentsline{toc}{subsection}{Acknowledgments}

This work was initiated during the 2017 ICERM semester program on Singularities and Waves in Incompressible Fluids. The work of RLP and YP is partially supported by the National Science Foundation under NSF grant DMS 1515400, the Simons Foundation under grant 395796 and the Center for Nonlinear Analysis (through NSF grant OISE-0967140).


\bigskip\bigskip
\addcontentsline{toc}{section}{References}
\bibliographystyle{abbrv}
\bibliography{biblio}
\bigskip\bigskip

\end{document}